# Stability Trend of Tilted Perovskites

Fazel Shojaei,[†‡] Wan-Jian Yin[†‡*]

[†]Soochow Institute for Energy and Materials Innovations (SIEMIS), College of Physics, Optoelectronics and Energy & Collaborative Innovation Center of Suzhou Nano Science and Technology, and

[‡]Key Laboratory of Advanced Carbon Materials and Wearable Energy Technologies of Jiangsu Province, Soochow University, Suzhou 215006, China

Abstract

Halide perovskites, with prototype cubic phase $ABX_3$, undergo various phase transitions accompanied by rigid rotations of corner-sharing $BX_6$ octahedra. Using first-principles density functional theory calculations, we have performed a comprehensive investigation of all the possible octahedral tilting in eighteen halide perovskites $ABX_3$ (A = Cs, Rb, K; B= Pb, Sn; X= I, Br, Cl) and found that the stabilization energies *i.e.* energy differences between cubic and the most stable tilted phases, are linearly correlated with tolerance factor *t*. Moreover, the tilt energies *i.e.* energy differences between cubic and various tilted phases, are linearly correlated with the change of atomic packing fractions ($\Delta\eta$), confirming the importance of atomic packing fraction as part of stability descriptor $(t+\mu)^\eta$, proposed in our previous work [JACS 139, 14905 (2017)]. We further demonstrate that $(t+\mu)^\eta$ remains the best stability descriptor for tilted perovskites among descriptor candidates of $\eta$, $\mu$, $t$, and $t+\mu$, extending previously proposed stability trend from cubic phases to tilted phases in general perovskites.

Table of Content (TOC):

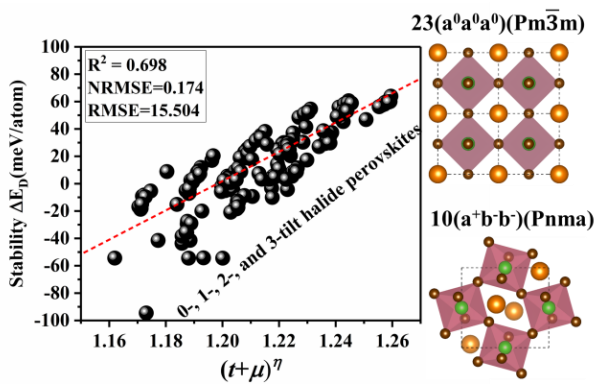

Solar cells based on halide perovskite materials of the form $ABX_3$ are currently the fastest growing photovoltaic technology in terms of research and development.[1-3] The idealized perovskite structure is cubic with space group of $Pm\bar{3}m$, comprising a highly flexible framework built up from chains of corner-sharing $[BX_6]$ octahedrons and A cations occupying the resulting holes with cuboctahedra symmetry. However, in real cases, cubic perovskite exhibits dynamical instabilities[4] and usually transforms to lower-symmetry phases as function of temperature and pressure, accompanying octahedral rotations about its symmetry axes ([100], [010], [001]) due to steric effects caused by ionic size mismatch.[5-10] For halide perovskites, literature review reveals that there are only three common phases extensively studied, namely, untilted cubic ($\alpha$-phase, $a^0a^0a^0$ in Glazer notation), tilted tetragonal ($\beta$-phase, $a^0a^0c^+$) and orthorhombic ($\gamma$-phase, $a^+b^-b^-$).[4,11-17] For example of $CsSnI_3$, the high-temperature cubic phase is observed above 440 K and it transforms to a tetragonal (P4/mbm) at 431 K, and further to an orthorhombic phase (Pnma) at 352 K.[5] Such kind of octahedral titling in halide perovskites can also be achieved by alloying with different size A-site cations.[11,18] To have a complete picture of perovskite tilt, Glazer developed a method for describing the octahedral tilting in perovskites,[19] which suggested a notation to describe any octahedral tilt as a combination of rotations about the three orthogonal symmetry axes of the octahedron. For example, $a^0b^+c^-$ represents a tilt system with two distinct rotation amplitudes: in-phase rotation about the [010] (b+) and out of phase rotation about [001] (c-) directions, which will be adopted in this letter for describing tilted perovskites. It initially showed that only *twenty three* different tilt systems can theoretically exist in perovskite structures. Nevertheless, in 1998, a group theory analysis of Glazer tilt systems led to only *fifteen* simple and unique tilt patterns.[20]

Although octahedral tilt is a common nature in perovskites, the relative stabilities and underlying trends of those tilted perovskites are rarely explored. Meanwhile, as a widely-accepted descriptor for perovskite stability, tolerance factor $t$, involving the ionic radius of A, B and X, is not able to distinguish the relative stabilities among different tilted phases of the same $ABX_3$ perovskite. Recently, we identified a stability descriptor for cubic perovskites, in which a strong linear correlation was observed between their decomposition energies ($\Delta H_D$) and the descriptor $(t+\mu)^\eta$, where $t$, $\mu$ and $\eta$ are the tolerance factor, the octahedral factor, and atomic packing fraction (AFP), respectively.[15] Although both $\mu$ and $t$ are only dependent on the ionic radius, $\eta$ is directly correlated with cell volumes, which may distinguish different phases. Therefore, it would be interesting to see if our previously found trend can extend to tilted perovskites.

In this letter, we performed first-principles density functional theory (DFT) calculations on fifteen possible tilted phases [Table 1] of eighteen halide perovskites $ABX_3$ with A = Cs, Rb, K; B= Pb, Sn; and X= I, Br, Cl. We have found that: *(i)* for all the eighteen perovskites, cubic phases are the most unstable phase and orthorhombic phase are the most stable phase; *(ii)* the stabilization energies, which are defined as the energy differences between cubic and most stable titled phases, are linearly correlated with tolerance factor $t$; *(iii)* the tilt energy, which are defined as the energy difference between cubic and various tilted phases, are linearly correlated with the change of atomic packing fractions ($\Delta\eta$); *(iv)* the stability descriptor $(t+\mu)^\eta$, which was previously proposed based on cubic phases can also apply to tilted perovskites, performing much better than conventional descriptor *t*.

For clarity and convenience, we used the same tilt number (TN) as those Glazer used in his original paper on the classification of different tilted perovskites.[19] Table 1 listed the notations of Glazer, Glazer's TN, the space group symmetries, and numbers of formula units in the primitive cell of the tilt systems. The crystal structures of five selected common phases are shown in Figure 1, with other ten phases shown in Figure S1 in Supporting Information. To distinguish eighteen kinds of ABX$_3$ perovskite compounds and fifteen different phases, henceforth, we call 'perovskite' as general name for all fifteen phases but with the same ABX$_3$ compositions and 'phase' as different tilted phase. To identify a phase, we combine TN, Glazer notation and the space group symmetries for clarity. For example, prototype cubic phases will be called 'tilt 23(a$^0$a$^0$a$^0$) (Pm$\bar{3}$m)'. We used SPuDS[21] and POTATO[22] software to construct initial crystal structures of tilted perovskites which were relaxed by subsequent DFT calculations. The DFT calculations are performed using the plane-wave pseudopotential approach within PBEsol functional[23] as implemented in Vienna ab initio simulation package (VASP).[24,25] The electron-ion interactions were described by the projector-augmented wave (PAW) method[26] with (n-1)$s^2$(n-1)$p^6$n$s^1$ for Cs/Rb/K, n$s^2$n$p^2$ for Pb/Sn, and n$s^2$n$p^6$ for I/Br/Cl as valance electrons. Both lattice constants and ions were relaxed in the direction of the Hellmann-Feynman force using the conjugate gradient method with an energy cut-off of 400 eV and *k*-spacing grid of 2π×0.015 Å$^{-1}$ until a stringent convergence criterion (0.001 eV/Å) was satisfied.

We define the *tilt energy* $\Delta E_{TN}^{tilt}$(ABX$_3$) = $E_{TN}^{tilt}$ (ABX$_3$) − $E_{23}^{tilt}$ (ABX$_3$), where $E_{TN}^{tilt}$ and $E_{23}^{tilt}$ are total energies of phases with tilted numbers TN and 23 respectively for ABX$_3$ perovskite. We recall in table 1 that the tilt number 23 phase is cubic. The tilt energies of eighteen perovskites with fifteen tilt phases are shown in Figure 2, with the numerical values given in Table S1. We find that for all perovskites, the orthorhombic γ-phase with tilt 10(a$^+$b$^-$b$^-$)(Pnma) is the ground state, while cubic α-phase with tilt 23(a$^0$a$^0$a$^0$) (Pm$\bar{3}$m) is the most unstable one. This finding is in line with the experimental observations that γ- and α-phases are the low- and high-temperature phases respectively for most halide perovskites including CsSnBr$_3$, CsSnI$_3$, CsPbCl$_3$, CsPbBr$_3$, and CsPbI$_3$.[5-7,27,28] Although the values of $\Delta E_{10}^{tilt}$(ABX$_3$) are the most negative for each perovskite, they span over a wide energy range from -4.66 meV/atom to -63.22 meV/atom for Sn-based perovskites ASnX$_3$ and from -17.48 meV/atom to -89.08 meV/atom for Pb-based perovskites APbX$_3$, demonstrating different tilting propensity for different perovskites. Since tilt 10(a$^+$b$^-$b$^-$)(Pnma) is found to be the ground state structure for all perovskites, $\Delta E_{10}^{tilt}$ (ABX$_3$) can be used to measure the tilting propensity for each perovskite. Therefore, we defined the value of $\Delta E_{10}^{tilt}$ (ABX$_3$) as the *stabilization energy* for perovskite ABX$_3$. More negative stabilization energy indicates higher tilting propensity. Our analysis shows that stabilization energy is almost insensitive to the chemical nature of halogen X, while it changes drastically with changing A and B cations. It can be seen that perovskites with either smaller A-cation or larger B-cation or both, have lower stabilization energy and thus higher propensity to tilt. Interestingly, the observed trend can be nicely explained by Goldschmidt's tolerance factor (*t*) which is an empirical descriptor to assess the structural stability and distortions in perovskites as shown in Figure 3.

The tolerance factor evaluates whether A-cation can fit within the holes created by the chains of [BX$_6$] octahedrons. An ideal perovskite compound adopts an untilted cubic structure with *t* =1. When the relative size of ions (A, B, and X) are not matched ideally, the tolerance factor deviates from *t* =1 due to the geometrical steric effects.[12,30] Meanwhile, when a *t* value falls between 0.9 and 1, it is still considered as a very good fit for perovskites, implying the possibility of cubic structure. Nevertheless, a *t* value between 0.71 and 0.9 implies the possible formation of orthorhombic or tetragonal structures with [BX$_6$] octahedral tilting.[29] In agreement with the tolerance factor predictions, all perovskite compounds we considered in this work which have *t* values in the range of $0.74 \leq t \leq 0.86$ exhibit tilted ground state structures. In Figure 3, it can be seen that the stabilization energy is almost linearly correlated with *t* which indicates that tolerance factor can be a good descriptor for the propensity to tilt in perovskites.

So far we have shown that there is a quantitative correlation between the stabilization energies and tolerance factors of the perovskites. Now, we want to take a closer look at the relative stabilities of different tilted phases of each perovskite. Since the tolerance factor *t* only depends on the chemical nature (ionic radii) of constituent ions of a perovskite, it is the same for all fifteen tilted phases for the same ABX$_3$ therefore cannot distinguish their stability differences. A suitable stability descriptor for tilted phases of perovskite is expected to be based on a traceable structural variable. As a structure-traceable factor within our recently proposed stability descriptor $(t+\mu)^\eta$, atomic packing fraction $\eta$ is defined as

$$\eta = \frac{\frac{4\pi}{3}[(r_A^3)+(r_B^3)+3(r_X^3)]}{V_{tilt}} \quad (1)$$

, where r$_A$, r$_B$, and r$_X$ are the ionic radius of A, B, and X ions and V$_{tilt}$ is the cell volume for tilted phase. It is apparent that when perovskites are tilted, their volumes shrink and the environments of A-cations change. In addition, the amount of shrinkage is different for different tilted phases. Therefore the volume and thus the APF may distinguish different tilted phases. Our calculations show that for each perovskite, tilted phases with larger $\eta$ values are more stable. In other words, perovskites energetically prefer tilts with dense atomic packing. These observations are in agreement with results reported by Woodward,[30] showing that the tilt with maximum number of short A-X interactions has the lowest energy and explaining the better performance of stability descriptor $(\mu+t)^\eta$ than $\mu+t$ by adding $\eta$ as a power index. Figure 4 shows tilt energy as function of $\eta$ for six bromide perovskites. An almost perfectly linear relationship between the tilt energy and $\eta$ is observed in all perovskites including iodide and chloride perovskites as shown in Figure S2, indicating that the APF is an important factor impacting the stability of tilted perovskites. It can be also seen from the Figure 4 that the maximum change of APF for each perovskite ($\eta_{10}^{ABX_3} - \eta_{23}^{ABX_3}$) increases from CsBBr$_3$ to KBBr$_3$ and from ASnBr$_3$ to APbBr$_3$. Exactly similar trends are also observed for chloride and iodide perovskites [Figure S2].

Another interesting finding is that for all the eighteen perovskites with fifteen tilted phases, although all the absolute APF values differ from each other, the relative APF values are linearly correlated with their tilted energies as shown in Figure 5(a), where the relative APF value for perovskite ABX$_3$ with tilt number TN is defined as $\Delta\eta_{TN}(ABX_3) = \eta_{10}(ABX_3) - \eta_{23}(ABX_3)$. Since the ionic radii are fixed in a certain perovskite ABX$_3$, the change of APF directly originated from the change of cell volume based on Eq. (1). Therefore, the $\Delta E_{TN}^{tilt}$-$\Delta\eta_{TN}$ correlation is in line with $\Delta E_{TN}^{tilt}$-$\Delta V_{TN}$ correlation as shown in Figure 5(b), where $\Delta V_{TN}$ is defined the fractional volume shrinkage of tilted TN phase compared to cubic phase *i.e.* $\Delta V_{TN} = (V_{23} - V_{TN})/V_{23}$, where $V_{TN}$ and $V_{23}$ are obtained volumes for the tilted TN phase and the cubic phase [tilt 23(a$^0$a$^0$a$^0$) (Pm$\bar{3}$m)], respectively.

Above, we have shown the general trends of stabilization energies on tolerance factor *t* and tilt energies on atomic packing fraction $\eta$ or fractional volume change $\Delta V_{TN}$. It would be interesting to investigate their thermodynamic stabilities and see whether previously proposed stability descriptor $(t+\mu)^\eta$, which already included the AFP, can still apply to tilted perovskites. The decomposition energies ($\Delta H_D$) of eighteen perovskites with fifteen phases are calculated and their plots dependent on *t*, $\mu$, $\eta$, *t*+$\mu$, and $(t+\mu)^\eta$ are shown in Figure 6. The accuracy rate of each descriptor to predict the relative stabilities among two tilt systems are also are also calculated according to the procedure presented in the supporting information of Ref. 15. We can see that: *(i)* Although $\Delta\eta$ is found to be a perfect stability descriptor for each individual perovskite, there exists no obvious correlation between $\Delta H_D$ and $\Delta\eta$ for collective perovskites, because the absolute $\eta$ value differ a lot for different perovskites; *(ii)* the linear correlation is even worse for $\mu$, *t* and $\mu$+*t* compared to the results of cubic phases in Ref. 15, since those ionic-radii dependent descriptors do not include information of structural distortion therefore are not able to describe the relative stabilities of tilted phases; *(iii)* in perfect agreement with our previous results, when we modify *t*+$\mu$ by adding $\eta$ as the power index, the linear trend is significantly improved which is manifested by the largest R$^2$ value obtained for the best linear fit (0.70 for $(t+\mu)^\eta$ *vs* 0.51 for *t*+$\mu$). As a descriptor, $(t+\mu)^\eta$ predicts the relative stabilities with an accuracy rate of 85%, which is much better than those calculated for *t* (74%), $\mu$ (52%), $\eta$ (41%), *t*+$\mu$ (75%). In short, we have shown that so far $(t+\mu)^\eta$ remain the best descriptor for the stability of halide perovskites including cubic and tilted phases.

In conclusion, we performed a comprehensive investigation of the impact of octahedral tilting on the stability of halide perovskites using first-principles density functional theory calculations. We have shown that all tilted phases are energetically more stable than their cubic untilted parents, whereby orthorhombic γ phase with tilt 10(a$^+$b$^-$b$^-$)(Pnma) is found to be the ground state. Several stability trends of tilted perovskites are identified: *(i)* the stabilization energies of different perovskites are linearly correlated with tolerance factors *t*; *(ii)* the tilt energies of different tilted phases are linearly correlated to the atomic packing fraction. We have further found that the stability descriptor $(t+\mu)^\eta$, which was recently proposed based on cubic phases, can be generalized to tilted perovskites, performing much better than $\eta$, $\mu$, *t* , and $\mu$+*t*. This work demonstrates direct evidences for the importance of atomic packing fraction and provide clear picture and trends to understand the stability of perovskites in their ideal and tilted phases.

**AUTHOR INFORMATION**

Corresponding author: Wan-Jian Yin, Email: wjyin@suda.edu.cn, Tel: +86-0512-67167457

**Notes**

The authors declare no competing final interest.

**ACKNOWLEDGMENT**

The authors acknowledge the funding support from National Natural Science Foundation of China (under Grant No. 51602211, No. 11674237), National Key Research and Development Program of China under grant No. 2016YFB0700700, Natural Science Foundation of Jiangsu Province of China (under Grant No. BK20160299), Jiangsu 'Double Talent' Program and Suzhou Key Laboratory for Advanced Carbon Materials and Wearable Energy Technologies, China. The work was carried out at National Supercomputer Center in Tianjin and the calculations were performed on TianHe-1(A).

**Supporting Information.** Chemical structures of less commonly observed tilted $ABX_3$ perovskites (Figure S1), tilting energies of chlorides and iodide perovskite as function of APF ($\eta$), numerical values of $\Delta E_{TN}^{tilt}(ABX_3)$ for Sn- and Pb-based (b) halide perovskites.

References

1. Green, M. A.; Bein, T. Photovoltaics: Perovskite Cells Charge Forward. *Nat. Mater.* **2015**, *14*, 559−561.
2. Stranks, S.; Snaith, H. J. Metal-halide Perovskites for Photovoltaic and Light-Emitting Devices. *Nat. Nanotechnol.* **2015**, *10*, 391−402.
3. Green, M. A.; Ho-Baillie, A.; Snaith, H. J. The Emergence of Perovskite Solar Cells. *Nat. Photonics* **2014**, *8*, 506−514.
4. Yang, R. X.; Skelton, J. M.; da Silva, E. L.; Frost, J. M.; Walsh, A. Spontaneous Octahedral Tilting in the Cubic Inorganic Cesium Halide Perovskites $CsSnX_3$ and $CsPbX_3$ (X = F, Cl, Br, I). *J. Phys. Chem. Lett.* **2017**, *8*, 4720−4726.
5. Chung, I.; Song, J. H.; Im, J.; Androulakis, J.; Malliakas, C. D.; Li, H.; Freeman, A. J.; Kenney, J. T.; Kanatzidis, M. G. $CsSnI_3$: Semiconductor or Metal? High Electrical Conductivity and Strong Near infrared Photoluminescence from a Single Material. High Hole Mobility and Phase-transitions. *J. Am. Chem. Soc.* **2012**, *134*, 8579− 8587.

6. Fabini, D. H.; Laurita, G.; Bechtel, J. S.; Stoumpos, C. C.; Evans, H. A.; Kontos, A. G.; Raptis, Y. S.; Falaras, P.; Van der Ven, A.; Kanatzidis, M. G. Dynamic Stereochemical Activity of the $Sn^{2+}$ Lone Pair in Perovskite $CsSnBr_3$. *J. Am. Chem. Soc.* **2016**, *138*, 11820−11832.
7. Trots, D.; Myagkota, S. High-temperature Structural Evolution of Cesium and Rubidium Triiodoplumbates. *J. Phys. Chem. Solids* **2008**, *69*, 2520−2526.
8. Baikie, T.; Fang, Y.; Kadro, J. M.; Schreyer, M.; Wei, F.; Mhaisalkar, S. G.; Graetzel, M.; White, T. J. Synthesis and Crystal Chemistry of the Hybrid Perovskite $(CH_3NH_3)PbI_3$ for Solid-State Sensitised Solar Cell Applications. *J. Mater. Chem. A* **2013**, *1*, 5628−5641.
9. Zhang, R.; Cai, W.; Bi, T.; Zarifi, N.; Terpstra, T.; Zhang, C.; Verdeny, Z. V.; Zurek, E.; Deemyad S. Effects of Nonhydrostatic Stress on Structural and Optoelectronic Properties of Methylammonium Lead Bromide Perovskite. *J. Phys. Chem. Lett.* **2017**, *8*, 3457–3465.
10. Postorino, P.; Malavasi, L. Pressure-Induced Effects in Organic−Inorganic Hybrid Perovskites. *J. Phys. Chem. Lett.* **2017**, *8*, 2613−2622.
11. Prasanna, R.; Gold-Parker, A.; Leijtens, T.; Conings, B.; Babayigit, A.; Boyen, H.-G.; Toney, M. F.; McGehee, M. D. Band Gap Tuning via Lattice Contraction and Octahedral Tilting in Perovskite Materials for Photovoltaics. *J. Am. Chem. Soc.* **2017**, *139*, 11117–11124.
12. Lee, J.-H.; Bristowe, N. C.; Lee, J. H.; Lee, S.-H.; Bristowe, P. D.; Cheetham, A. K.; Hyun Jang, H. M. Resolving the Physical Origin of Octahedral Tilting in Halide Perovskites *Chem. Mater.* **2016**, *28*, 4259–4266.
13. Marronnier, A.; Lee, H.; Geffroy, B.; Even, J.; Bonnassieux, Y.; Roma, G. Structural Instabilities Related to Highly Anharmonic Phonons in Halide Perovskites. *J. Phys. Chem. Lett.* **2017**, *8*, 2659−2665.
14. Young, J.; Rondinelli, J. M. Octahedral Rotation Preferences in Perovskite Iodides and Bromides. *J. Phys. Chem. Lett.* **2016**, *7*, 918–922.
15. Sun, Q.; Yin, W.-j. Thermodynamic Stability Trend of Cubic Perovskites. *J. Am. Chem. Soc.* **2017**, *139*, 14905-14908.
16. Sun, S.; Deng, Z.; Wu, Y.; Wei, F.; Isikgor, F. H.; Federico Brivio, F.; Michael W. Gaultois, M. W.; Jianyong Ouyang, J.; Paul D. Bristowe, P. D.; Anthony K. Cheetham, A. K.; Gregor Kieslich, G. Variable Temperature and High-Pressure Crystal Chemistry of Perovskite Formamidinium Lead Iodide: a Single Crystal X-ray Diffraction and Computational Study. *Chem. Commun.* **2017**, *53*, 7537-7540.
17. Meloni, S.; Palermo, G.; Ashari-Astani, N.; Grätzel, M.; Rothlisberger, U. Valence and Conduction Band Tuning in Halide Perovskites for Solar Cell Applications. *J. Mater. Chem. A* **2016**, *4*, 15997-16002.
18. Linaburg, M. R.; McClure, E. T.; Majher, J. D.; Woodward, P. M. $Cs_{1−x}Rb_xPbCl_3$ and $Cs_{1−x}Rb_xPbBr_3$ Solid Solutions: Understanding Octahedral Tilting in Lead Halide Perovskites. *Chem. Mater.* **2017**, *29*, 3507–3514.
19. Glazer, A. M. The Classification of Tilted Octahedra in Perovskites. *Acta. Cryst.* **1972**, *B28*, 3384.
20. Howard, C. J.; Stokes, H. T. Group Theoretical Analysis of Octahedral Tilting in Perovskites. *Acta Cryst.* **1998**, *B54*, 782-789.

21. M. W. Lufaso, M. W.; Woodward, P. M. Prediction of the Crystal Structures of Perovskites Using the Software Program SPuDS. *Acta Cryst.* **2001**, *B57*, 725-738.
22. http://www.unf.edu/~michael.lufaso/spuds/potato.html
23. Perdew, J. P.; Ruzsinszky, A.; Csonka, G. I.; Vydrov, O. A.; Scuseria, G. E.; Constantin, L. A.; Zhou, X.; Burke, K. Restoring the Density-Gradient Expansion for Exchange in Solids and Surfaces. *Phys. Rev. Lett.* **2008**, *100*, 136406.
24. Kresse, G.; Hafner, J. Ab initio Molecular Dynamics for Liquid Metals. *Phys. Rev. B: Condens. Matter Mater. Phys.* **1993**, *47*, 558−561.
25. Kresse, G.; J. Furthmüller, J. Efficient Iterative Schemes for Ab Initio Total-Energy Calculations Using a Plane-Wave Basis Set. *Phys. Rev. B: Condens. Matter Mater. Phys.* **1996**, *54*, 11169−11186.
26. Kresse, G.; D. Joubert, D. From Ultrasoft Pseudopotentials to the Projector Augmented-Wave Method. *Phys. Rev. B: Condens. Matter Mater. Phys.* **1999**, *59*, 1758−1775.
27. Fujii, Y.; Hoshino, S.; Yamada, Y.; Shirane, G. Neutron-Scattering Study on Phase Transitions of $CsPbCl_3$. *Phys. Rev. B* **1974**, *9*, 4549−4559.
28. Hirotsu, S.; Harada, J.; Iizumi, M.; Gesi, K. Structural Phase Transitions in $CsPbBr_3$. *J. Phys. Soc. Jpn.* **1974**, *37*, 1393−1398.
29. Ju, M.-G.; Dai, J.; Ma, L.; Zeng, X. C. Lead-Free Mixed Tin and Germanium Perovskites for Photovoltaic Application. *J. Am. Chem. Soc.* **2017**, *139*, 8038–8043.
30. Woodward, P. M. Octahedral Tilting in Perovskites. II. Structure Stabilizing Forces. *Acta Cryst.* **1997**, *B53*, 44-66.

**Table 1** Notations of Glazer, Glazer's tilt numbers, the space group symmetries, and numbers of ABX$_3$ formula units in the simulation cell (N) of fifteen possible tilts.

|  | Tilt class | Tilts | Tilt number | space-group | N |
|---|---|---|---|---|---|
| zero-tilt | 000 | $(a^0a^0a^0)$ | 23 | $Pn\bar{3}m$(#221) | 1 |
| One-tilt | 00- | $(a^0a^0c^-)$ | 22 | $I4/mcm$(#140) | 4 |
|  | 00+ | $(a^0a^0c^+)$ | 21 | $P4/mbm$(#127) | 2 |
| two-tilt | 0-- | $(a^0b^-b^-)$ | 20 | $Imma$(#74) | 4 |
|  |  | $(a^0b^-c^-)$ | 19 | $C2/m$(#12) | 4 |
|  | 0+- | $(a^0b^+c^-)$ | 17 | $Cmcm$(#63) | 8 |
|  | 0++ | $(a^0b^+b^+)$ | 16 | $I4/mmm$(#139) | 8 |
| three-tilt | --- | $(a^-a^-a^-)$ | 14 | $R\bar{3}c$(#167) | 6 |
|  |  | $(a^-b^-b^-)$ | 13 | $C2/c$(#15) | 8 |
|  |  | $(a^-b^-c^-)$ | 12 | $P1$(#2) | 8 |
|  | +-- | $(a^+b^-b^-)$ | 10 | $Pnma$(#62) | 4 |
|  |  | $(a^+b^-c^-)$ | 8 | $P2_1/m$(#11) | 8 |
|  | ++- | $(a^+a^+c^-)$ | 5 | $P4_1/nmc$(#137) | 8 |
|  | +++ | $(a^+a^+a^+)$ | 3 | $Im\bar{3}$(#204) | 8 |
|  |  | $(a^+b^+c^+)$ | 1 | $Immm$(#71) | 8 |

**Figure 1.** Crystal structures of five commonly observed tilted phases for ABX$_3$ halide perovskites with tilts (a) cubic 23($a^0a^0a^0$), (b) tetragonal 21($a^0a^0c^+$), (c) tetragonal 22($a^0a^0c^-$), (d) cubic 3($a^+a^+a^+$), and (d) orthorhombic 10($a^+b^-b^-$). Each phase can be described as a combination of rotations about the three orthogonal symmetry axes of the octahedral. Here, signs 0, +, and – denote no tilting, in-phase rotation and out-of-phase rotation about that axis. Light brown, green, and dark brown colors represent A, B, and X ions.

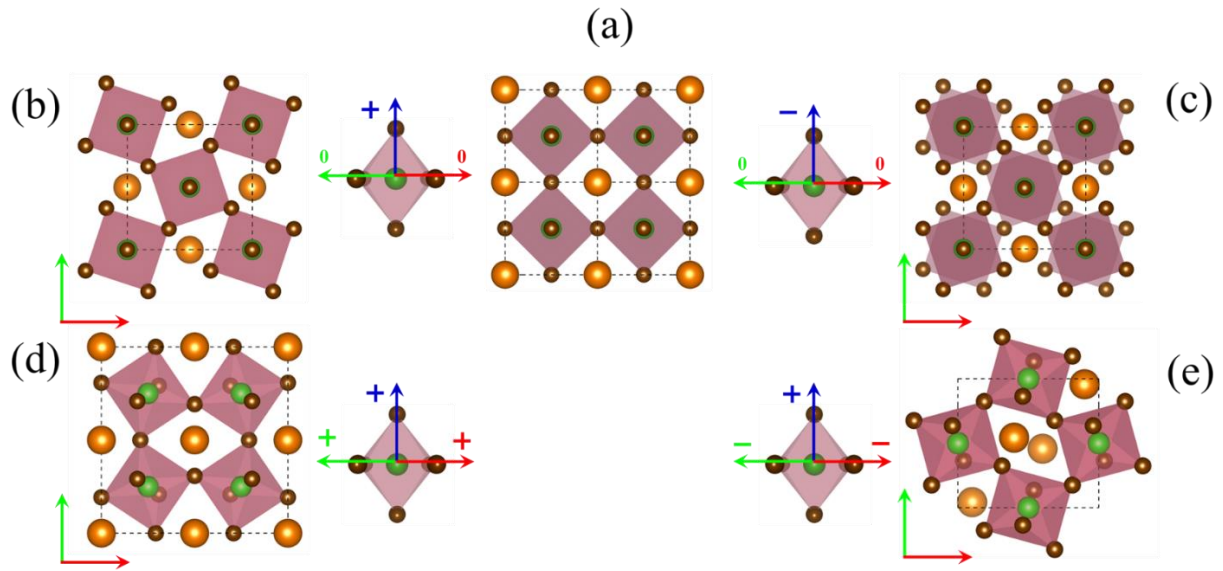

**Figure 2.** The tilted energies of eighteen perovskites with fifteen tilted phases of each perovskite relative to the corresponding tilt $23(a^0a^0a^0)$ ($\Delta E_{TN}(ABX_3) = E_{TN}(ABX_3) - E_{23}(ABX_3)$).

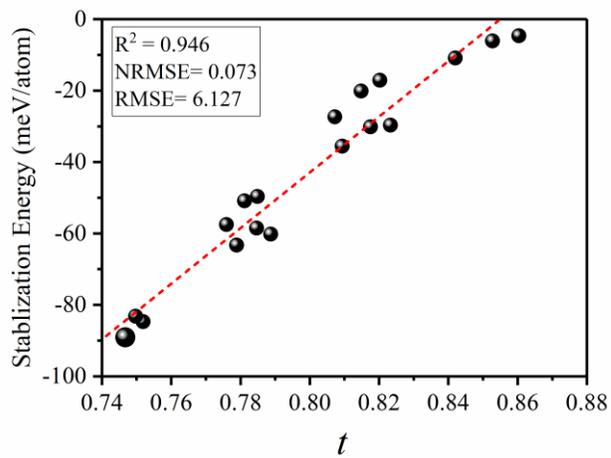

**Figure 3.** Calculated stabilization energies as function of the tolerance factor ($t$) for eighteen halide perovskites $ABX_3$ [A = Cs, Rb, K; B= Pb, Sn; X= I, Br, Cl]. The coefficient of determination ($R^2$), normalized root mean square error (NRMSE), and root mean square error (RMSE) for the best fit line are also shown. $R^2$ and NRMSE are dimensionless, while RMSE is in meV/atom unit.

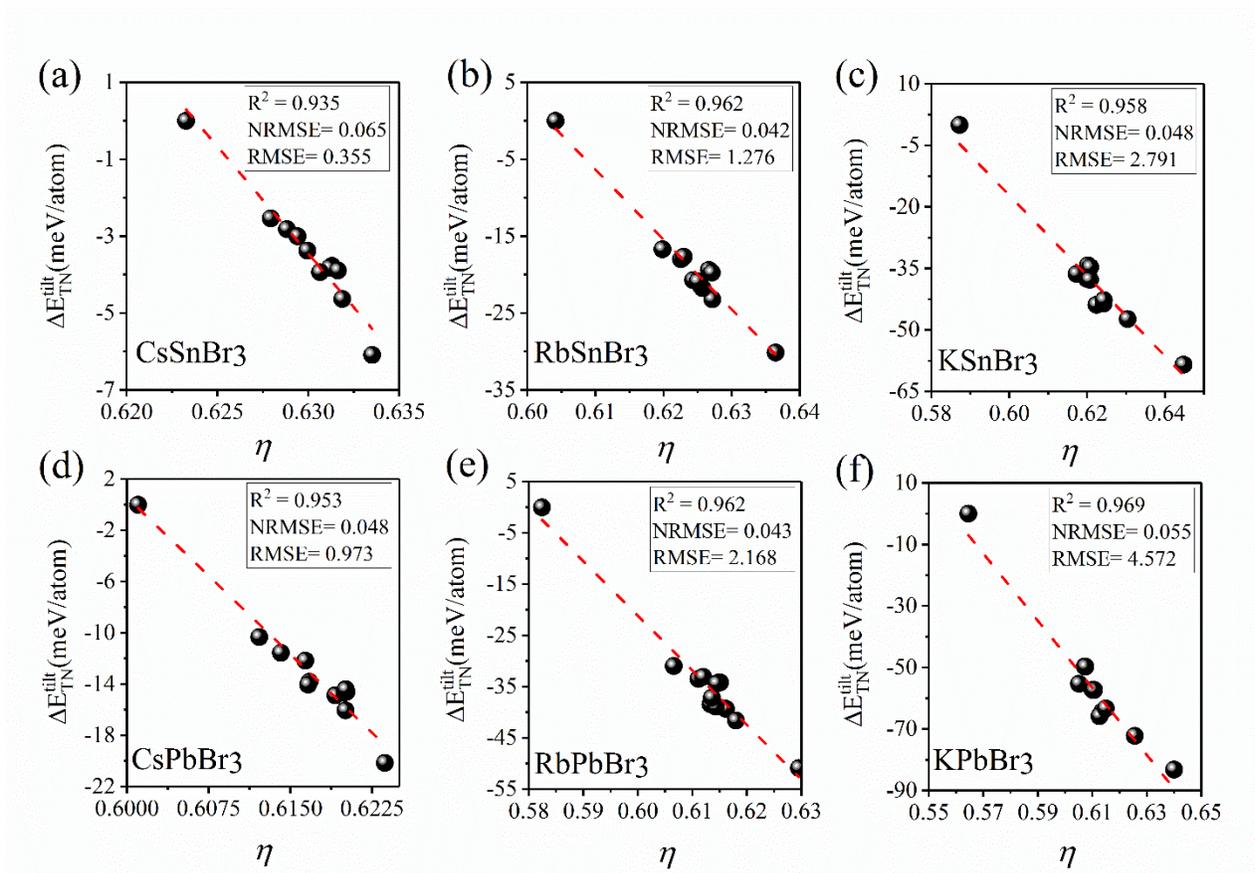

**Figure 4.** Energies of different tilts of each bromide perovskite relative to the corresponding tilt 23($a^0a^0a^0$) as function of APF ($\eta$). Note that the scales of y-axis are different for each plot.

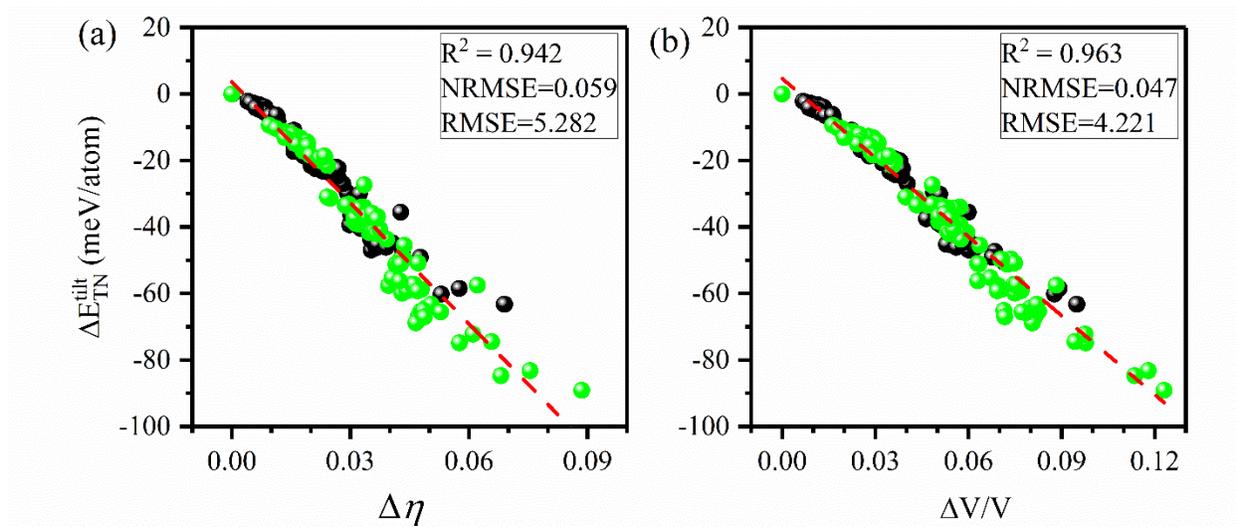

**Figure 5.** Tilted energies of eighteen halide perovskites $ABX_3$ [A = Cs, Rb, K; B= Pb, Sn; X= I, Br, Cl] with fifteen tilted phases as functions of (a) relative APF values and (b) fractional volume shrinkage. Black and green colors represent Sn- and Pb-based perovskites, respectively.

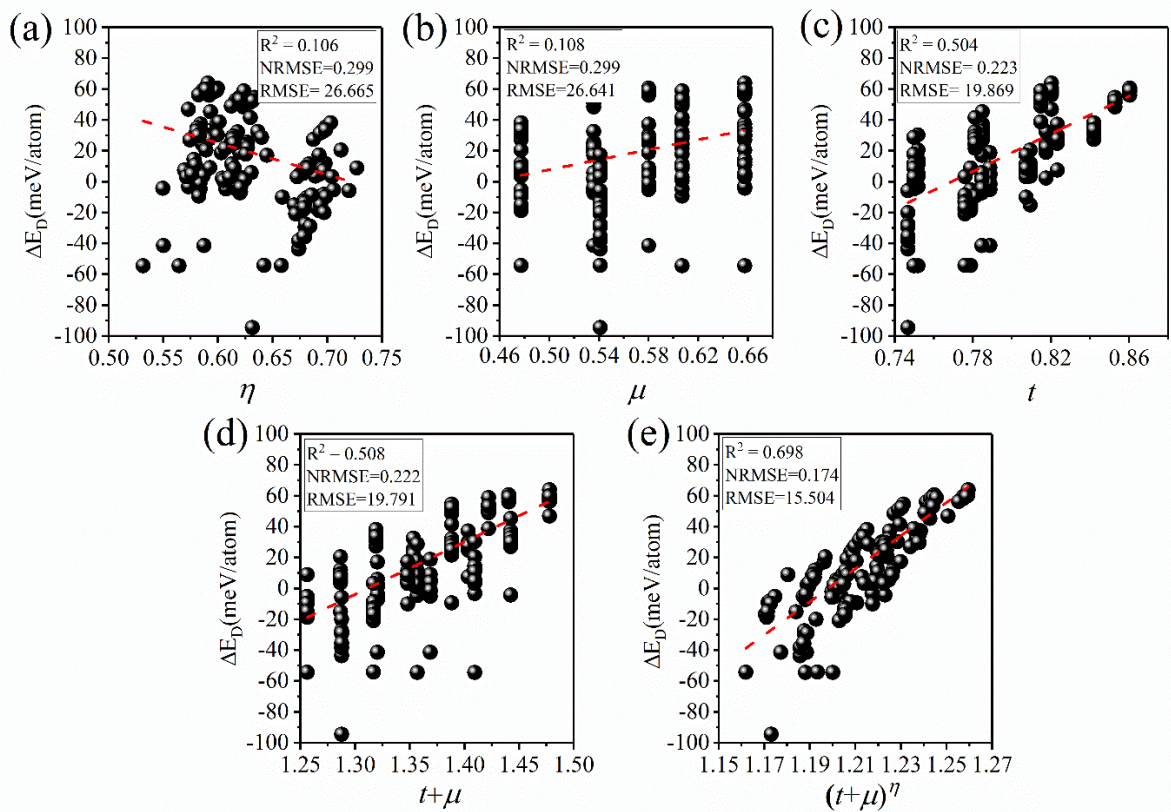

**Figure 6.** Decomposition energies (ΔH$_D$) of halide perovskites dependent on $\eta$ (a), $\mu$ (b), $t$ (c), $t+\mu$ (d), and $(t+\mu)^\eta$ (e).

Supporting information

# Stability Trend of Tilted Perovskites

Fazel Shojaei,[†‡] Wan-jian Yin[†‡*]

[†]Soochow Institute for Energy and Materials Innovations (SIEMIS), College of Physics, Optoelectronics and Energy & Collaborative Innovation Center of Suzhou Nano Science and Technology, and

[‡]Key Laboratory of Advanced Carbon Materials and Wearable Energy Technologies of Jiangsu Province, Soochow University, Suzhou 215006, China

*Corresponding Author: wjyin@suda.edu.cn

**Figure S1.** Chemical structures of other ten less commonly observed ABX$_3$ perovskites. Light brown, green, and dark brown colors represent A, B, and X ions.

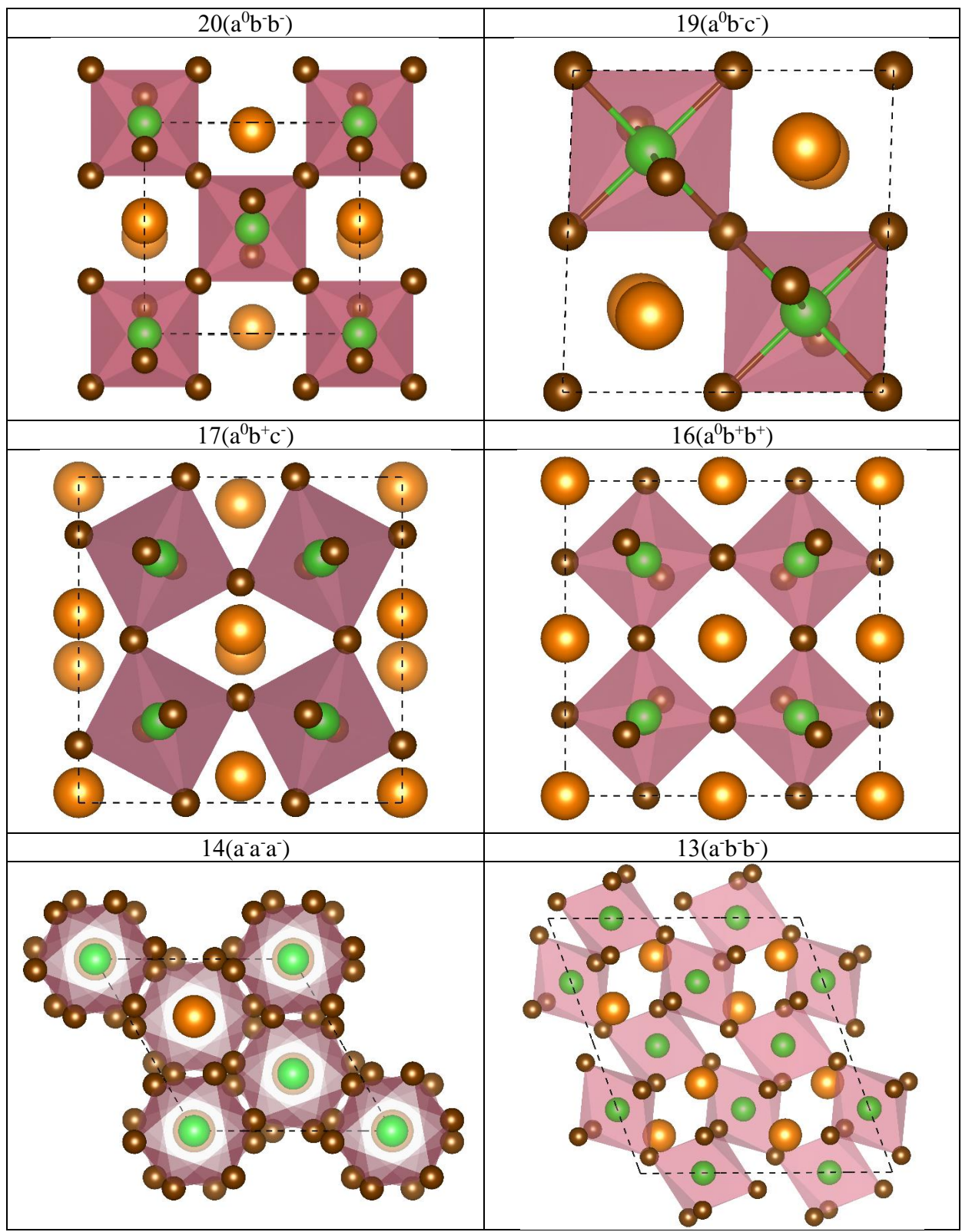

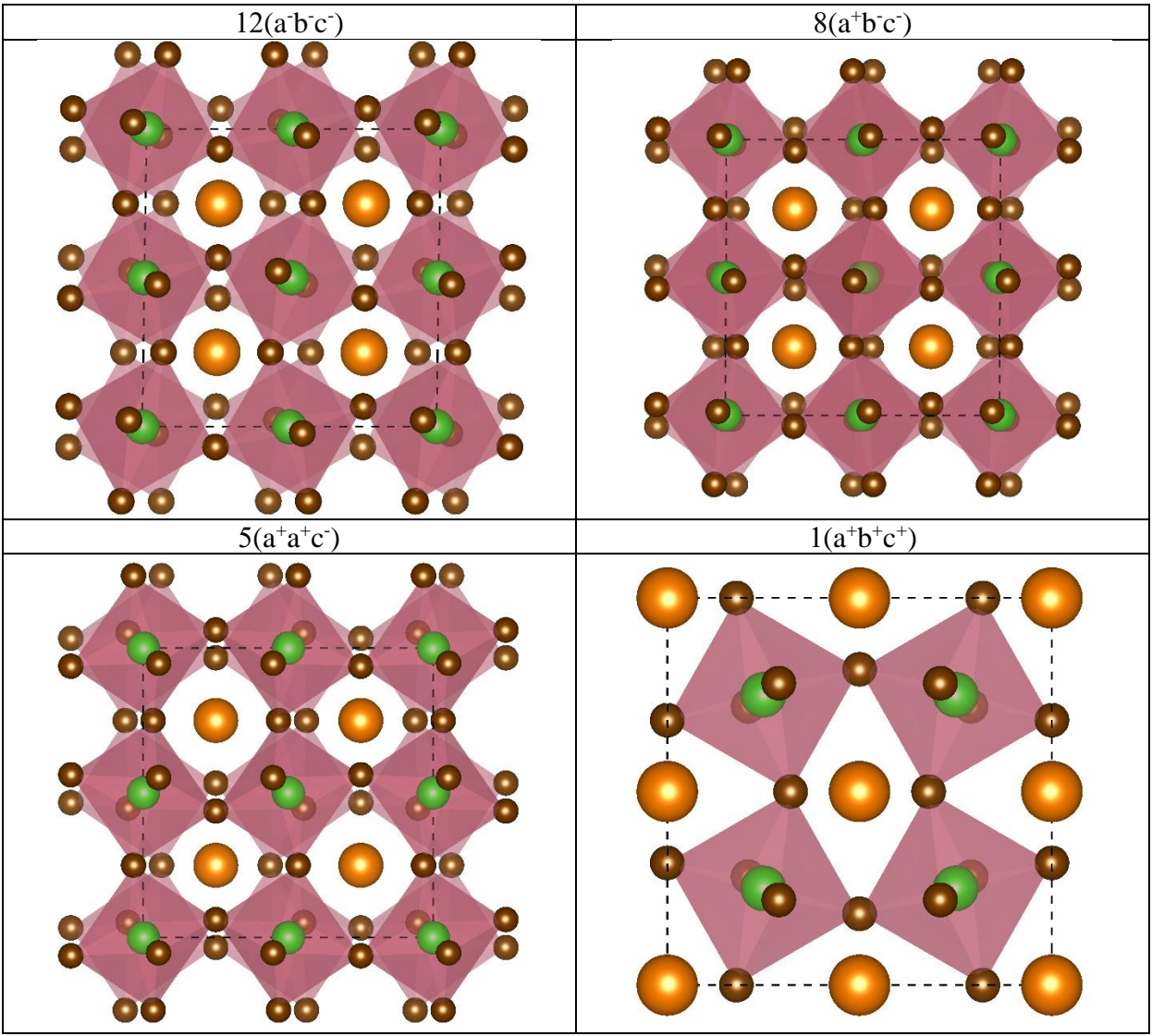

**Figure S2.** Energies of different tilts of each bromide perovskite relative to the corresponding tilt 23($a^0a^0a^0$) ($Pm\bar{3}m$) as function of APF ($\eta$). The coefficient of determination ($R^2$), normalized root mean square error (NRMSE), and root mean square error (RMSE) for the best fit line are also shown for each one. $R^2$ and NRMSE are dimensionless, while RMSE is in meV/atom unit.

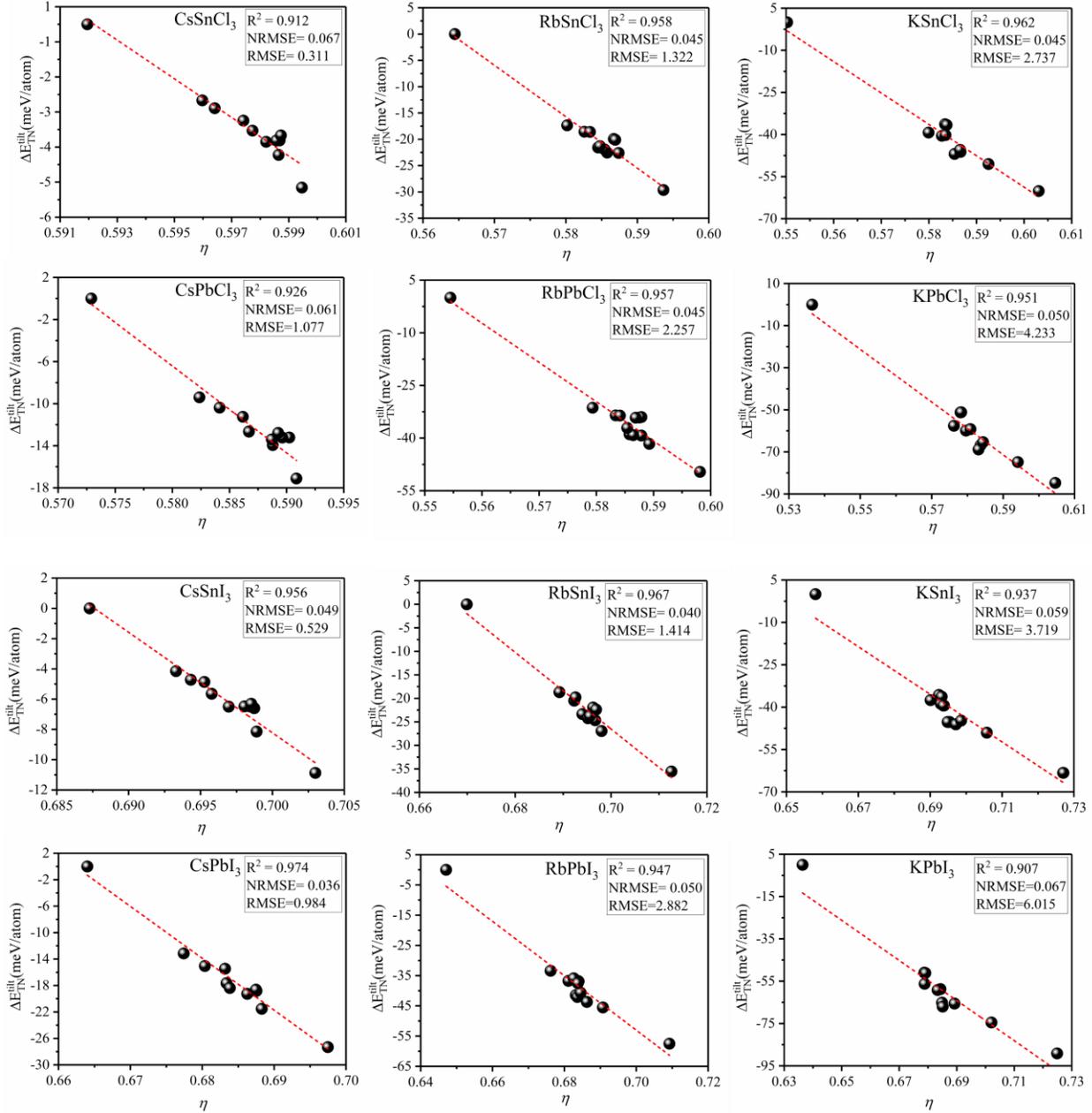

**Table S1**. The numerical values of $\Delta E^{tilt}_{TN}(ABX_3)$ calculated for Sn- (a) and Pb-based (b) halide perovskites. All energy values are given in meV/atom unit.

(a)

| Tilt number | Cl | | | Br | | | I | | |
|---|---|---|---|---|---|---|---|---|---|
| | Cs | Rb | K | Cs | Rb | K | Cs | Rb | K |
| 23 | 0 | 0 | 0 | 0 | 0 | 0 | 0 | 0 | 0 |
| 22 | -3.310 | -19.941 | -36.323 | -3.774 | -19.393 | -34.258 | -6.482 | -21.944 | -35.662 |
| 21 | -3.167 | -20.103 | -36.580 | -3.889 | -19.745 | -34.697 | -6.599 | -22.388 | -36.274 |
| 20 | -3.030 | -21.560 | -46.938 | -3.371 | -20.718 | -43.844 | -5.644 | -23.355 | -45.269 |
| 19 | -3.314 | -21.942 | -46.938 | -3.824 | -21.356 | -43.844 | -6.480 | -24.241 | -45.260 |
| 17 | -3.723 | -22.597 | -46.210 | -4.630 | -23.209 | -43.530 | -8.145 | -26.959 | -46.106 |
| 16 | -2.392 | -18.562 | -40.211 | -2.815 | -17.998 | -37.749 | -4.719 | -20.495 | -39.452 |
| 14 | -2.745 | -18.593 | -40.463 | -3.002 | -17.688 | -37.559 | -4.861 | -19.774 | -38.823 |
| 13 | -3.030 | -22.539 | -50.520 | -3.372 | -21.758 | -47.325 | -5.644 | -24.668 | -49.055 |
| 12 | -3.314 | -21.942 | -60.156 | -3.824 | -21.356 | -58.474 | -6.480 | -24.241 | -63.263 |
| 10 | -4.654 | -29.647 | -60.156 | -6.092 | -30.147 | -58.474 | -10.877 | -35.549 | -63.263 |
| 8 | -4.654 | -29.647 | -60.156 | -6.092 | -30.147 | -58.474 | -10.877 | -35.549 | -63.263 |
| 5 | -3.350 | -21.293 | -45.464 | -3.934 | -20.888 | -42.683 | -6.499 | -23.785 | -44.764 |
| 3 | -2.168 | -17.354 | -39.313 | -2.534 | -16.690 | -36.348 | -4.148 | -18.682 | -37.518 |
| 1 | -3.167 | -20.103 | -40.213 | -3.889 | -19.745 | -37.743 | -6.306 | -22.388 | -39.455 |

(b)

| Tilt number | Cl | | | Br | | | I | | |
|---|---|---|---|---|---|---|---|---|---|
| | Cs | Rb | K | Cs | Rb | K | Cs | Rb | K |
| 23 | 0 | 0 | 0 | 0 | 0 | 0 | 0 | | 0 |
| 22 | -13.215 | -34.148 | -51.223 | -14.603 | -34.112 | -49.685 | -18.805 | -36.778 | -50.936 |
| 21 | -12.778 | -34.015 | -51.176 | -14.399 | -34.158 | -49.757 | -18.679 | -36.893 | -51.099 |
| 20 | -12.638 | -38.919 | -68.839 | -13.812 | -38.416 | -65.810 | -17.650 | -41.441 | -66.999 |
| 19 | -13.394 | -39.154 | -68.835 | -14.873 | -38.871 | -65.830 | -19.257 | -42.038 | -67.005 |
| 17 | -13.935 | -39.277 | -66.894 | -16.045 | -39.303 | -64.397 | -21.522 | -43.697 | -65.129 |
| 16 | -10.382 | -33.544 | -59.190 | -11.565 | -33.488 | -57.259 | -15.040 | -36.724 | -59.184 |
| 14 | -11.248 | -33.589 | -59.913 | -12.170 | -33.094 | -57.224 | -15.457 | -35.863 | -58.747 |
| 13 | -12.662 | -41.634 | -74.829 | -14.043 | -41.574 | -72.217 | -18.375 | -45.519 | -74.509 |
| 12 | -13.394 | -39.146 | -68.823 | -14.881 | -38.904 | -65.838 | -19.274 | -42.058 | -67.011 |
| 10 | -17.108 | -49.626 | -84.731 | -20.147 | -50.857 | -83.217 | -27.319 | -57.495 | -89.126 |
| 8 | -17.107 | -49.626 | -84.731 | -20.147 | -50.857 | -83.217 | -27.319 | -57.495 | -89.126 |
| 5 | -13.223 | -37.105 | -65.379 | -14.647 | -37.122 | -63.262 | -18.853 | -40.662 | -65.570 |
| 3 | -9.396 | -31.364 | -57.607 | -10.336 | -30.960 | -55.259 | -13.143 | -33.372 | -56.227 |
| 1 | -12.789 | -34.204 | -59.207 | -14.435 | -34.365 | -57.291 | -18.628 | -37.494 | -59.182 |